\documentclass[prb,aps,amsmath,amssymb,superscriptaddress,showpacs]{revtex4}
\usepackage[latin1]{inputenc}
\usepackage{graphicx}
\usepackage[all]{xy}
\usepackage[amsmath,thmmarks]{ntheorem}

\newcommand*\colvec[1]{\begin{pmatrix}#1\end{pmatrix}}
 
\def\ext{{\rm ext}}

\def\A{{\mathcal A}}

\def\K{{\mathcal K}}
\def\D{{\mathcal D}}

\def\B{{\mathcal B}}

\def\CC{{\mathbb C}}
\def\HH{{\mathbb H}}
\def\NN{{\mathbb N}}

\def\ad{{\mathrm ad}}
\def\Ad{{\mathrm Ad}}

\def\End{\mbox{\rm End}}

\def\bra{\langle}
\def\ket{\rangle}

\def\tr{\mbox{\rm Tr}}

\def\bea{\begin{eqnarray}}
\def\eea{\end{eqnarray}}

\def\be{\begin{equation}}
\def\ee{\end{equation}}

\newtheorem{theorem}{Theorem}

\newtheorem{lemma}{Lemma}
 
\newtheorem{propo}{Proposition}
\theoremstyle{nonumberplain}
\theorembodyfont{\normalfont}
\theoremseparator{:}
\theoremsymbol{$\P$}
\newtheorem{demo}{Proof}

\newenvironment{rem}[1][{}]{\smallbreak \noindent  {\bf Remark #1}\small }

\begin{document}
\title{Extensions of the noncommutative Standard Model and the weak order one condition}

\author{Fabien Besnard}
 \affiliation{P{\^o}le de recherche M.L. Paris,
 EPF, Graduate school of engineering,
 3~bis rue Lakanal,
 F-92330 Sceaux, France}

\begin{abstract}
In the derivation of the Standard Model from the axioms of Noncommutative Geometry, the scalar sector is given by a finite Dirac operator which has to satisfy the so-called \emph{first-order condition}. However, the general solution to this constraint still has unphysical terms which must be fine-tuned to zero. Moreover, the first-order condition generally does not survive in  extensions to models with gauge groups larger that $U(1)\times SU(2)\times SU(3)$. In this paper we show that  in the $U(1)_{\rm B-L}$-extension one can implement a weaker form of the first-order condition which we argue is necessary in order for Noncommutative Gauge Theory to make sense at all, and that this condition reduce the amount of fine-tuning to the off-diagonal terms in the Yukawa mass matrices for the leptons and quarks. We also show that this condition eliminates the Majorana mass terms for right-handed neutrinos when it is applied to the Pati-Salam model.
\end{abstract}
\maketitle

\section{Introduction}
The Standard Model of particle physics has many features which seem arbitrary. One must choose    a gauge group and its representation  among a plethora of possibilities and add a scalar sector which has \emph{a priori} no reason to limit itself to a single Higgs doublet. The Lagrangian is constrained by the theoretical principle of gauge invariance and the practical necessity of renormalizability, but the CP violating terms of QCD must still be fine-tuned away, giving rise to the so-called strong CP problem. A theory with the ability to produce the Standard Model or some of its extensions but with a lesser degree of  arbitrariness would thus be very welcome. Noncommutative Geometry is such a theory.

It is well-known that particle models coming from the Noncommutative Geometry (NCG) approach are subjects to severe theoretical constraints, and it is no small feat that the Standard Model emerges as one of the few possibilities. To begin with, in such models one has to start with a finite-dimensional algebra\footnote{There exist attempts to constrain the choice of the algebra \cite{connesmarcolli}, but we will not consider them in this paper.}, which is more constraining that the choice of a Lie group. Moreover there are far less representations available for algebras than for groups. Nonetheless, the gauge  and   fermionic fields are ultimately inferred from experiment, just as in the usual case, and this fixes the algebra $\A_F$ and its representation. This algebra is then tensorized with the functions on spacetime to produce a so-called \emph{almost-commutative} manifold,   a kind of Kaluza-Klein product manifold except that the compact part is replaced with a finite noncommutative space. On such an object the metric is incarnated in a Dirac operator which is a sum of two terms: one coming from the canonical Dirac operator on the manifold, and the other being a matrix  $D_F$ acting on the finite-dimensional space of fermions. The presence of $D_F$ naturally generates the scalar sector of the theory, that is the field content as well as the Yukawa couplings. This is already an improvement over usual model building, especially because  $D_F$ has to fulfil a large number of theoretical constraints: it must be selfadjoint, commute with charge conjugation and anti-commute with chirality. It is often asked that it also satisfies the so-called \emph{first-order condition} (C1, to be recalled below), which is inferred from the commutative case when it is equivalent to the Dirac being a first-order differential operator. In the case of the Standard Model, these conditions remove many fields and terms in the Lagrangian, but still leave open the possibility for some unwanted scalars. This is why an even more constraining \emph{``second-order condition''} has been under investigation \cite{BF1,BF2,bbb,dds}. This condition is once again inferred from the commutative case, where it manifests itself as the graded-commutation of $1$-forms. However,   it is quite cumbersome to implement the second-order condition in a general almost-commutative manifold. In this note we explore the opposite direction which is to \emph{weaken} the order 1 condition. This weakening seems to be necessary in order to extend the Standard Model \cite{ccvs}. However, without the order 1 condition the guiding principles of NCG are not guaranteed  to make sense anymore. To explain our concern, we have to distinguish between \emph{Noncommutative Gauge Theory} (NGT), introduced by Connes and Lott \cite{conneslott}, and the Spectral NCG of Chamseddine and Connes \cite{cc96}. Both are defined on a \emph{spectral triple}, the replacement for a Riemannian manifold in NCG, but the former has a bosonic action which is of Yang-Mills type and defined only on the gauge and scalar degrees of freedom, while the latter uses the spectral action which  also includes gravity terms. However, the  spectral triple includes a fixed background metric. In order to promote Spectral NCG to  a  full-fledged noncommutative \emph{noncommutative Kaluza-Klein theory},  one has to specify a replacement for the background differential manifold on which the metric  may vary. A recent proposal is the notion of \emph{algebraic background} \cite{algbackpart1}. In this framework it can be established that the first-order condition ensures 3 key properties: 
\begin{enumerate}
\item gauge-transformations are symmetries of the background,
\item the space of gauge and Higgs fields is a subset of the total configuration space of all the Dirac operators compatible with the differential structure,
\item this subset is gauge-invariant.
\end{enumerate}
These conditions can be seen as the weakest which make NGT a consistent submodel of noncommutative Kaluza-Klein theory. We call them weak C1, C1', and C1''. They are weaker than C1 even if taken together \cite{blstandalone}. In this paper we will explore their consequences for the B-L extension of the SM defined in ref. \onlinecite{blstandalone}. This  extension is a NGT defined on a background $\B^\ext$ which contains the SM background $\B^{\rm SM}$. We will suppose that 
\begin{itemize}
\item $\B^{\rm SM}$ satisfies C1 as usual,
\item $\B^\ext$ satisfies weak C1, C1' and C1'',
\end{itemize}
and show that this sets to zero exactly the same elements in the Dirac operator as the second-order condition did for the Standard Model. As a result, the scalar sector of the extended model   consist, in the most general case, of two Higgs doublets and a complex singlet responsible for the breaking of B-L symmetry and giving a Majorana mass to the right-handed neutrinos. The appearance of a second Higgs doublet is problematic since, as we explicitly show, its vaccum state is not electrically neutral, thus giving a mass to the photon. The faulty terms in the Dirac operator can be set to zero, making the second Higgs vanish, but giving rise to a fine-tuning problem in NGT not unlike the strong CP problem. In spite of this, it is remarkable that NGT comes quite close to uniquely providing a well-motivated and long-studied abelian extension of the SM with all its intricacies from first principles. 

In this paper we put the emphasis on NGT in Lorentzian signature, partly because Spectral NCG has only been defined in Euclidean signature to date, partly because in the absence of C1 the bosonic degrees of freedom are represented in a technically more involved way in Spectral NCG \cite{ccvs}. However, in the B-L extension we are mostly interested in in this paper, these complications disappear and the conclusion remains.

In the whole paper the symbols $\bar A, A^\dagger$ and $A^\times$ stand for the complex conjugate, Hilbert adjoint and Krein adjoint of $A$, respectively.  The word ``adjoint'' will always refer to the Krein adjoint unless otherwise specified.

\section{Noncommutative gauge theory}
In this section we recall the notions of NGT that we will need in this paper. For more details see ref. \onlinecite{blstandalone}. We suppose that spectral triples are familiar to the reader. For a definition, see ref. \onlinecite{connesmarcolli}.

An (even, real) \emph{algebraic background} (AB) is an (even, real) spectral triple deprived of its Dirac operator but augmented with a bimodule of $1$-forms. It is hence a tuple $\B=(\A,\K,\pi,\chi,J,\Omega^1)$ where $\A$ is a $*$-algebra, $\K$ a Krein space, $\pi$ a representation of $\A$ on $\K$, $\chi$ the chirality operator, $J$ the real structure anti-linear operator, and $\Omega^1\subset\End(\K)$ an odd $\A$-bimodule. For any operator $T$ on $\K$ we define its \emph{opposite} $T^o$ by the formula:
\be 
T^o:=JT^\times J^{-1}.
\ee
We  require  the so-called \emph{order zero condition} (C0) to hold:
\be 
[\pi(a)^o,\pi(b)]=0,\forall a,b\in \A.
\ee
A background is \emph{almost-commutative} (AC) when it is the (graded) tensor product $\B_M\hat\otimes \B_F$ with $\B_M$ the canonical AB over a manifold $M$ (see ref. \onlinecite{algbackpart1} for a definition) and $\B_F$ is a finite-dimensional background (the algebra $\A_F$ and the Krein space $\K_F$ are both finite-dimensional).

A \emph{compatible Dirac operator} for $\B$ is an operator which turns $(\B,D)$ into a spectral triple such that $\Omega^1_D\subset\Omega^1$, where $\Omega^1_D$ is the bimodule of $1$-forms $\sum_i \pi(a_i)[D,\pi(b_i)]$. When $\Omega^1_D=\Omega^1$ we say that $D$ is \emph{regular}. The configuration space $\D_\B$ of $\B$ is by definition the vector space of compatible Dirac operators. An automorphism (or symmetry) of an AB is a Krein-unitary operator $U$ such that $U\pi(\A)U^{-1}=\pi(A)$, $U\Omega^1U^{-1}=\Omega^1$, $UJ=JU$ and $U\chi=\chi U$. Automorphisms automatically preserve the configuration space.

What we call NGT in this paper (and should not be confused with gauge theory defined on a deformation of Minkowski space) is the theory devised by Connes and Lott\cite{conneslott}, extended to general signature for the base manifold in ref. \onlinecite{Elsner-99} and to general signature  in the finite part as well as to the presence of opposite forms in ref. \onlinecite{blstandalone}. We only review the general ideas below and refer to the these papers for more information. NGT is defined on an AC spectral triple with underlying background $\B=\B_M\hat\otimes\B_F$  and    Dirac operator 
\be 
D=D_g\hat\otimes 1+1\hat\otimes D_F
\ee
where $D_g$ is the canonical Dirac operator associated to a fixed background metric on $M$ and $D_F$ is a fixed  regular Dirac operator on $\B_F$.  Its bosonic variable is a selfadjoint 1-form $\omega$. With such a 1-form we build a \emph{fluctuated} Dirac operator
\be
D_\omega:=D+\omega+\omega^o\label{flucdir}
\ee
We write 
\begin{equation}
{\cal F}:=\{\omega+\omega^o|\omega\in\Omega^1\}
\end{equation}
and call it the \emph{space of fluctuations}. The action of NGT is the generalized Connes-Lott action:
\be
S_b(\omega)=\int_M \tr(P(\rho(\omega+\omega^o))^2) {\rm vol}_g\label{CLaction}
\ee
where $\rho$ is the curvature of $\omega+\omega^o$ and $P$ is a projection operator which we do not need to describe here. The power and beauty of this approach is that the variable $\omega$ contains all the gauge \emph{and} Higgs degrees of freedom through the decomposition
\be 
\Omega^1=\Omega_M^1\hat\otimes \A_F+\A_M\hat\otimes\Omega_F^1\label{decform}.
\ee
Moreover, $S_b$ unifies the Yang-Mills and Higgs terms (respectively in the curvature of the smooth and finite part of $\omega$) and predicts all the coupling constants and coefficients of the quartic potential at unification scale. It is also  automatically gauge-invariant, with the gauge transformations defined by
\be 
\Upsilon(u):=\pi(u)(\pi(u)^{-1})^o=\pi(u)J\pi(u)J^{-1},\mbox{ with }u\in\A\mbox{ s.t. }uu^\dagger=1.
\ee
Note that a unitary element of the algebra $\A$ is a map from $M$ to the unitary group of the finite-dimensional algebra $\A_F$.

It is clear that in order for NGT to make sense, 3 conditions must be met:
\begin{enumerate}
\item\label{wc1} gauge-transformations must be symmetries of the background,
\item\label{wc1p} the space of fluctuated Dirac operators must be a subset of the configuration space,
\item\label{wc1s} this subset must be stable under gauge transformations.
\end{enumerate}
These 3 conditions are all ensured by the following single axiom, which is called the \emph{first-order condition} (C1):
\be 
[\pi(a)^o,\Omega^1]=0, \forall a\in \A.
\ee
This axiom is motivated by the case of a canonical spectral triple over a manifold, where it holds precisely because the Dirac operator is a first-order differential operator. However, one needs to go beyond the first-order condition as soon as the Standard Model is extended. This is true for the Pati-Salam extension \cite{ccvs}, but also for the simpler B-L extension \cite{blstandalone}.  It has been shown\cite{pertsemigroup} that one should add a non-linear term $\omega_s$ to \eqref{flucdir} in order to obtain a gauge-invariant space even without C1. More precisely, if $\omega=\sum \pi(a_i)[D,\pi(b_i)]$ let us define
\be 
\omega_s=\sum J\pi(a_i)[\omega^o,\pi(b_i)]J^{-1},
\ee
and the space of \emph{perturbations} ${\cal P}=\{\omega+\omega^o+\omega_s|\omega\in\Omega^1\}$. Then it was shown \cite{pertsemigroup} that the space of perturbed Dirac operators $D+{\cal P}$ is stable under gauge transformations. However, we cannot use a perturbed Dirac operator in   \eqref{CLaction} since the curvature of a perturbation does not have an obvious meaning. We thus want to continue to use fluctuations and so have no option but to ask  for conditions \ref{wc1}, \ref{wc1p} and \ref{wc1s} separately. We call them, in order, the weak C1, C1' and C1'' conditions and explore their consequences in the next section.

\section{The weak $C_1,C_1'$ and $C_1''$ conditions}
Let us first put these conditions in  algebraic form. The weak order 1 condition is asking that for all $u$ in the unitary group of $\A$, $\Upsilon(u)$ is a background automorphism. It is immediately seen that $\Upsilon(u)$ is unitary and commutes with both $J$ and $\chi$. It stabilizes $\pi(\A)$ thanks to C0. Hence we see that weak C1 boils down to  the stabilization of $\Omega^1$, and   using  C0  and the fact that $\Omega^1$ is an $\A$-bimodule, this is  equivalent to the following condition: for all unitary $u$ one has:
\be 
\pi(u)^o\Omega^1(\pi(u)^o)^{-1}\subset\Omega^1\mbox{ (WC1) }\label{WC1}
\ee
Let $D_\omega=D+\omega+\omega^o$ be a fluctuated Dirac. Since $\omega$ is selfadjoint and odd, it is clear that $D_\omega$ is also selfadjoint and odd, and it is easy to see that it also commutes with $J$. The only thing to check is thus that $\Omega^1_{D_\omega}\subset \Omega^1$. This is true iff for all $a\in \A$, $[D_\omega,\pi(a)]\in\Omega^1$. Since $D$ is regular and $\Omega^1$ is a bimodule, we see that WC1' is equivalent to
\be 
[\omega^o,\pi(\A)]\subset \Omega^1\mbox{ (WC1') }
\ee
for all selfadjoint $\omega\in\Omega^1$. 

In order to put WC1'' in algebraic form let us introduce some notations. The action of a gauge transformation on a fluctuated Dirac $D_\omega$ is
\bea
\Upsilon(u)D_\omega \Upsilon(u)^{-1}&=&D+u(u^{-1})^o(\omega+\omega^o)u^ou^{-1}+u(u^{-1})^o[D,u^ou^{-1}]\cr
&:=&D+(\omega+\omega^o)^{\Upsilon(u)}\label{gtfluct}
\eea
where the last equality defines $(\omega+\omega^o)^{\Upsilon(u)}$ and we have suppressed the representation $\pi$ for the sake of clarity. Condition WC1'' requires that 

\be 
(\omega+\omega^o)^{\Upsilon(u)}\in {\cal F}\mbox{ (WC1'') }
\ee
for all unitary $u$. The next lemma is useful reformulation under WC1.

\begin{lemma}\label{icaution} If WC1 holds, then WC1'' is equivalent to
\be 
u\beta^ou^{-1}+\beta\in{\cal F},
\ee
where $\beta=u[D,u^{-1}]$ and $u$ is any unitary element of $\A$.
\end{lemma}
{\small\begin{demo}
We compute $(\omega+\omega^o)^{\Upsilon(u)}$ using   \eqref{gtfluct}. We obtain by C0:
\be
(\omega+(\omega)^o)^{\Upsilon(u)}=(u^{-1})^o\alpha u^o+((u^{-1})^o\alpha u^o)^o+u(u^{-1})^o[D,u^ou^{-1}],\label{rouge}
\ee
where $\alpha=u\omega u^{-1}$ is a 1-form. Now by WC1   $(u^{-1})^o\alpha u^o$ is a 1-form, so that the first two terms of \eqref{rouge} form a fluctuation, and WC1'' is equivalent to $u(u^{-1})^o[D,u^ou^{-1}]\in{\cal F}$ for all unitary $u$. We can clean up this expression by writing
\bea
u(u^{-1})^o[D,u^ou^{-1}]&=&u(u^{-1})^o([D,u^o]u^{-1}+u^o[D,u^{-1}])\cr
&=&u(u^{-1})^o[D,u^o]u^{-1}+u[D,u^{-1}]\label{eq41}
\eea
Let $\beta=u[D,u^{-1}]$. Then using $[D,uu^{-1}]=0$ one gets $\beta=-[D,u]u^{-1}$  and $\beta^o=(u^{-1})^o[D,u^o]$. Hence  \eqref{eq41} yields
\bea
u(u^{-1})^o[D,u^ou^{-1}]&=&u\beta^ou^{-1}+\beta
\eea
\end{demo}}

We can focus on the finite part of the background thanks to the following result:

\begin{propo}
Let $\B=\B_M\hat\otimes\B_F$ be an AC background and $D=D_M\hat\otimes 1+1\hat\otimes D_F$, where $D_M$ and $D_F$ are regular Dirac for $\B_M$ and $\B_F$ respectively. Then $(\B,D)$ satisfies weak C1, C1' and C1'' iff  $(\B_F,D_F)$ does.
\end{propo}
{\small\begin{demo}
A unitary element of $\A=\A_M\otimes \A_F$ is a smooth map $x\mapsto u(x)$ from $M$ to the unitary group of $\A_F$, and a 1-form of $\B$ can always be decomposed as $\omega=\omega'+\omega''$, where $\omega'$ is a 1-form over $M$ with values in $\A_F$ and $\omega''$ is a function on $M$ with values in $\Omega^1_F$ according to \eqref{decform}.

Suppose WC1, WC1' and WC1'' holds for the finite part. Then at every $x$ one has $u(x)^o\omega''(x)(u(x)^o)^{-1})\in\Omega^1_F$, so that $u^o\omega''(u^o)^{-1}\in\A_M\hat\otimes\Omega^1_F\subset \Omega^1$. On the other hand
 $u(x)^o\omega'(x)(u(x)^o)^{-1}=\omega'(x)$ by C0. Thus $\omega'$ commutes with $u^o$. Hence $\B$ satisfies WC1.
 
Consider pure tensor products $\omega'=\omega_M\hat\otimes a$ and $\omega''=f\hat\otimes \omega_F$, with $f\in\A_M$ and $a\in\A_F$. Then for any pure tensor $g\hat\otimes b$, $g\in\A_M, b\in\A_F$, one has
\bea
[\omega_M^o\hat\otimes a^o,g\hat\otimes b]&=&\omega_M^og\hat\otimes [a^o,b],\mbox{ by C1 on }\B_M\cr
&=&0,\mbox{ by C0 on }\B_F.
\eea
and
\bea
[f^o\hat\otimes \omega_F^o,g\hat\otimes b]&=&fg\hat\otimes[\omega_F^o,b]\cr
&\in&\A_M\hat\otimes \Omega^1_F,\mbox{ by WC1' on }\B_F.
\eea
Hence WC1' holds on $\B$ by linearity.

Now we  prove that WC1'' holds using lemma \ref{icaution}. We can write $\beta$ as the sum of its smooth part $\beta_M$ and finite part $\beta_F$ thanks to the decomposition of $D$:
\bea
\beta&=&u[D_M\hat\otimes 1+1\hat\otimes D_F,u^{-1}]\cr
&=&u[D_M\hat\otimes 1,u^{-1}]+u[1\hat\otimes D_F,u^{-1}]\cr
&:=&\beta_M+\beta_F
\eea
Now we have already observed that an element of $\Omega^1_M\hat\otimes \A_F$ like $\beta_M$ commutes with elements of $\A^o$, hence by taking opposites we have $[\beta_M^o,u]=0$. Thus
\be
u\beta^ou^{-1}+\beta=\beta_M^o+\beta_M+u\beta_F^ou^{-1}+\beta_F,
\ee
and we are reduced to proving that $u\beta_F^ou^{-1}+\beta_F$ is a fluctuation. Consider a point $x\in M$. Then since  $\B_F$ satisfies WC1 and WC1''  we obtain by lemma \ref{icaution} that $u(x)\beta_F^o(x)u^{-1}(x)+\beta_F(x)$ is a (finite) fluctuation, and this proves that $u\beta_F^ou^{-1}+\beta_F$ is of the form $\gamma+\gamma^o$ with $\gamma\in\A_M\hat\otimes\Omega^1_F$.

Conversely, if $(\B,D)$ satisfies WC1, WC1' and WC1'' then it is easy to prove by considering  constant functions and 1-forms that the finite part also does.
\end{demo}}

In the rest of this section we consider a finite-dimensional background $\B_F$. We first prove that WC1 can be put in infinitesimal form.

\begin{propo} Weak C1 is equivalent to:
\be 
[\pi(a)^o,\Omega^1_F]\subset\Omega^1_F, \label{derweakC1}
\ee
for every anti-selfadjoint $a\in\A_F$.
\end{propo}
{\small\begin{demo}
One goes from  \eqref{WC1} to  \eqref{derweakC1} by derivation. To show the converse, observe that \eqref{derweakC1} ensures that $\Omega^1_F$ is stable by ${\rm ad}_{\pi(a)^o}^n$ for all $n\in\NN$. Using $e^{{\rm ad}_X}={\rm Ad}_{e^X}$,  we obtain \eqref{WC1} for every $u$ of the form $e^a$. Now the exponential map to a Lie group from its Lie algebra is surjective when the group is compact and connected, which is the case of the unitary group of $\A_F$ (which is a direct sum of matrix algebras). Hence every $u$ is of the form $e^a$.
\end{demo}}

We now show that WC1'' is in fact a redundant condition. We start with a lemma.

\begin{lemma}\label{flucstab} If WC1 holds, the fluctuation space ${\cal F}$ is stable under $\ad_{a-a^o}$ for every anti-selfadjoint $a\in\A_F$.
\end{lemma}
{\small\begin{demo}
This follows from 
\bea
[a-a^o,\omega+\omega^o]&=&[a,\omega]-[a^o,\omega^o]-([a^o,\omega]-[a,\omega^o])\cr
&=&[a,\omega]+[a,\omega]^o-([a^o,\omega]+[a^o,\omega]^o).
\eea
Indeed, $[a,\omega]\in\Omega^1_F$ since it is a bimodule, and $[a^o,\omega]\in\Omega^1_F$ from WC1. Moreover, since $a$ is anti-selfadjoint and $\omega$ selfadjoint, $[a,\omega]$ and $[a^o,\omega]$ are both selfadjoint.
\end{demo}}

\begin{theorem} Condition WC1 implies condition WC1''  .
\end{theorem}
{\small\begin{demo}
Let $u=e^a$ with $a$ anti-selfadjoint. Since $a$ and $a^o$ commute, we have $\Upsilon(u)=e^ae^{(-a)^o}=e^{a-a^o}$. Hence
\bea
(\omega+\omega^o)^{\Upsilon(u)}&=&\Ad_{\Upsilon(u)}(D_\omega)-D\cr
&=&\Ad_{\exp(a-a^o)}(D_\omega)-D\cr
&=&e^{\ad_{a-a^o}}(D_\omega)-D\cr
&=&\sum_{k=1}^\infty \frac{1}{k!}\ad_{a-a^o}^k(D_\omega)+D_\omega-D\cr
&:=&E+\omega+\omega^o
\eea
We see that $(\omega+\omega^o)^{\Upsilon(u)}\in {\cal F}$ iff $E$ does. To show this, observe first that $\ad_{a-a^o}(D_\omega)=\ad_{a-a^o}(D)+\ad_{a-a^o}(\omega+\omega^o)=-([D,a]+[D,a]^o)+\ad_{a-a^o}(\omega+\omega^o)$. The first term is in ${\cal F}$ since $[D,a]$ is a selfadjoint 1-form and the second term is also in ${\cal F}$ thanks to lemma \ref{flucstab}. By induction we have $\ad_{a-a^o}^k(D_\omega)\in{\cal F}$ for every $k\ge 1$, so that $E\in{\cal F}$.
\end{demo}}

By taking opposites we see that weak C1' entails in particular that $[\pi(a)^o,\omega]$ is a $1$-form when $\omega$ is self-adjoint and $a$ anti-selfadjoint. However we cannot conclude that weak C1' entails weak C1 since the latter is not restricted to self-adjoint $1$-forms.

\section{The Standard Model background}\label{SMback}
The SM background is $\B_{\rm SM}=\B_M\hat\otimes \B_F$ where   $\B_M$ is the canonical background over a four-dimensional manifold\footnote{The manifold has to satisfy some properties which will not matter here. See ref. \onlinecite{algbackpart1} for details.} $M$, and $\B_F=(\A_F,\K_F,\ldots,\Omega^1_F)$ is the following finite background:

\begin{itemize}
\item $\A_F=\CC\oplus\HH\oplus M_3(\CC)$,
\item  The   Krein space is 
\begin{eqnarray}
\K_F&=&\K_R\oplus \K_L \oplus \K_{\bar R}\oplus\K_{\bar L}\label{KF},
\end{eqnarray}
where each $\K_i$ is  isomorphic to
\begin{eqnarray}
\K_0 &:=& \CC^2\otimes\CC^4\otimes \CC^3.\label{K0}
\end{eqnarray}
The $\CC^2$ factor represents weak isospin and $\CC^3$ is the generation space\footnote{When we need to distinguish the $\CC^3$ of quark colors from the generation space we write them $\CC^3_c$ and $\CC^3_g$ respectively. Note that  there can be any number of generations in what follows.}. The $\CC^4$ factor will often be split as
\be 
\CC^4=\CC_l\oplus \CC_q\label{lqdec}
\ee
where $\CC_l=\CC$ is the ``leptonicity space''  and $\CC_q=\CC^3$ is the quark color space. A block-dagonal matrix in the decomposition \eqref{KF} will be written as
\be 
[A,B,C,D]
\ee
where $A,B,C,D$ are operators on $\K_0$.
\item The representation is 
\begin{equation}
\pi_F(\lambda,q,a)=[q_\lambda\otimes 1_4,q\otimes 1_4,1_2\otimes(\lambda\oplus a), 1_2\otimes(\lambda\oplus a)]\otimes 1_3\label{SMrep}
\end{equation} 
where $q_\lambda=\colvec{\lambda&0\cr 0&\bar\lambda}$ and quaternions are seen as matrices of the form $\colvec{\alpha&\beta\cr -\bar\beta& \bar\alpha}$. Moreover $\lambda\oplus a=\colvec{\lambda&0\cr 0&a}$ written in the decomposition \eqref{lqdec}.
\item  The chirality operator is
\begin{equation}
\chi_F=[1,-1,-1,1]\label{SMchirality}
\end{equation}
\item  The Krein product on $\K_F$ is $(\psi_1,\psi_2)_F=\psi_1^\dagger\eta_F\psi_2$, where $\dagger$ is the Hilbert adjoint for the canonical scalar product on $\K_F$ and the fundamentaly symmetry $\eta_F=\chi_F$.
\item The real structure is
\begin{equation}
J_F=\colvec{0&0&-1&0\cr 0&0&0&-1\cr 1&0&0&0\cr 0&1&0&0 } \circ c.c.\label{JF}
\end{equation}
where c.c. means complex conjugation.
\item The bimodule $\Omega^1_F$ is left undetermined for the moment.
\end{itemize}
The logic of our approach is the following. We take $\A_F$, $\K_F$ and $\pi_F$ for granted : they yield the gauge and fermionic sector of the theory which we take as an experimental input. The operators $\chi_F,\eta_F$ and $J_F$ are then uniquely determined up to a global phases \cite{thesenadir,barrettunique} by the non-triviality of the fermionic action and the possibility to solve fermion quadrupling by Barrett's conditions. The question we seek to answer is: what freedom is remaining in the choice of $\Omega^1_F$, or in physically equivalent terms, what are the constraints on the scalar sector of the theory ?

Since we need a regular Dirac operator $D_F$ for NGT, we can as well explore the constraints on $D_F$. From $D_F=D_F^\times$, $D_FJ_F=J_FD_F$ and $D_F\chi_F=-\chi_F D_F$ we see that this Dirac must have the form
\be
D_F=\colvec{0&-Y^\dagger &-M^\dagger &0\cr Y&0&0&-B^\dagger\cr M&0&0&-Y^T\cr 0&B&\bar Y&0}\label{finitedirac}
\ee

with $M=M^T$ and $B=B^T$. We will need below to write the matrices $Y,B,M$ in blocks in the lepton/quark decomposition of $\CC^4$ as $M=\colvec{M_{ll}&M_{lq}\cr M_{ql}&M_{qq}}$, $B=\colvec{B_{ll}&B_{lq}\cr B_{ql}&B_{qq}}$,  and $Y=\colvec{Y_{ll}&Y_{lq}\cr Y_{ql}&Y_{qq}}$. The blocks of these matrices are elements of $\End(\CC^2\otimes \CC^3_g)$. If we choose to decompose them with respect to $\CC^2$ the sub-blocks will be indexed $uu,ud,du,uu$. For instance we can write $M_{ll}=\colvec{M_{ll}^{uu}&M_{ll}^{ud}\cr M_{ll}^{du}&M_{ll}^{dd}}$ where each blocks now acts only on $\CC^3_g$.

We want $\B_{\rm SM}$ to satisfy the first-order condition.  The implications of  C1 on the Dirac operator are well-known \cite{connesmarcolli,bbb}, hence we only quickly recall them here. C1 is equivalent to the four following equations:
\begin{enumerate}
\item\label{equa1} $[Y q_\lambda- q Y, \lambda'\oplus m']=0$,
\item\label{equa2} $(B q-\lambda\oplus m B)(\lambda'\oplus m')= {q'}(B q-(\lambda\oplus m)B)$,
\item\label{equa3} $[Y^T,\lambda\oplus m] {q}'= {q_\lambda}'[Y^T,\lambda\oplus m]$,
\item\label{equa4} $(M^\dagger(\lambda\oplus m)- {q_\lambda}M^\dagger) {q_{\lambda'}}=\lambda'\oplus m'(M^\dagger (\lambda\oplus m)- q_\lambda M^\dagger)$,
\end{enumerate}
for all $\lambda,\lambda'\in\CC,q,q'\in \HH,m,m'\in M_3(\CC)$. Note that we have suppressed obvious tensor products: $q$ means $q\otimes 1_4\otimes 1_3$ and $\lambda\oplus m$ means $1_2\otimes(\lambda\oplus m)\otimes 1_4$.

Equation \ref{equa2} immediately gives $B=0$. Equation \ref{equa1} yields that $Y=\colvec{Y_l&0\cr 0&Y_q}$, that is, $Y$ is diagonal in the lepton/quark decomposition of $\CC^4$. Equation \ref{equa3} shows that $Y_q$ acts trivially in the color space ($Y_q$ only acts on the doublet and generations indices).  Finally  equation \ref{equa4} entails that $M=\colvec{M_{ll}&M_{ql}^T\cr M_{ql}&0}$, with $M_{ll}^{dd}=0$, $M_{ql}^{du}=0$ and $M_{ql}^{dd}=0$.

The remaining independent   submatrices of $M$ are thus $M_{ll}^{uu}$, $M_{ll}^{ud}$, $M_{ql}^{uu}$ and $M_{ql}^{ud}$. If we ask for the second-order condition to hold (either in its original formulation\cite{BF1} or in graded form \cite{bbb}) then only $M_{ll}^{uu}$ remains. Here we will arrive at the same conclusion by enforcing WC1  on the extended B-L background.

\section{The B-L extended background}
It has been shown  \cite{algbackpart2} that gauged B-L symmetries are automorphisms of the SM background, so that the Standard Model gauge group has to be extended by $U(1)_{B-L}$ in order to be consistent with the algebraic background point of view.  We therefore define the extended background $\B_F^\ext$, where the only changes with respect to $\B_F$ are in the algebra and 1-forms which become respectively $\A_F^\ext=\CC\oplus\CC\oplus\HH\oplus M_3(\CC)$ and $(\Omega^1_D)^\ext$, which is generated by commutators of $D_F$ with elements of $\A_F^\ext$. The representation is
\be 
\pi_F^\ext(\lambda,\mu,q,m)=[q_\lambda,q,\mu\oplus m,\mu\oplus m]
\ee
where, once again, obvious tensor products with identity matrices are understood. We just write this representation $\pi$ below when no confusion may arise.

We define 
\begin{equation}
\omega_{Y}=\colvec{0&0&0&0\cr Y&0&0&0\cr 0&0&0&0 \cr 0&0&0&0},\mbox{ and }\omega_{M}=\colvec{0&0&0&0\cr 0&0&0&0\cr M&0&0&0\cr 0&0&0&0}.
\end{equation}

\begin{lemma}
The family $G=\{\omega_{Y},\omega_{M},\omega_Y^\times,\omega_M^\times\}$ generates $(\Omega^1_D)^\ext$ as a $\A_F^\ext$-bimodule.
\end{lemma}
{\small\begin{demo}
The bimodule $(\Omega^1_D)^\ext$ is   generated as a vector space by elements of the form $\pi(a')[D,\pi(a)]=$
{\small
\begin{equation}
\colvec{ 0&-  q_{\lambda'}Y^\dagger q+  q_{\lambda'\lambda}Y^\dagger &-  q_{\lambda'}M^\dagger(\mu\oplus m)+  q_{\lambda'\lambda}M^\dagger &0\cr 
 q'Y  q_\lambda-  q'  qY&0&0&0\cr
(\mu'\oplus m')M  q_\lambda-(\mu\mu'\oplus m'm)M&0&0&0\cr
 0&0&0&0}\label{formeomega1}
\end{equation}
}
Now let $V$ be the bimodule spanned by $G$. Clearly $(\Omega^1_D)^\ext\subset V$, and to show the converse it suffices to realize the elements of $G$ as   elements of the form \eqref{formeomega1}. To obtain 
\begin{itemize}
\item $\omega_Y$, we set $q'$ and $\lambda$ to $1$ and $q,\lambda',\mu,m,\mu',m'$ to $0$,
\item $\omega_M$, we set $\mu',m',\lambda$ to $1$ and the others to $0$.
\end{itemize}
The  proof is completed by taking adjoints.
\end{demo}}

\section{The weak order $1$ condition on $\B_F^\ext$}\label{sectionWC1}
We require $\B_F^\ext$ to satisfy WC1. First we observe that thanks to C0 we have for all $a,b,c\in\A_F^\ext$ and $\omega\in(\Omega_F^1)^\ext$,
\bea
[\pi(a)\omega \pi(b),\pi(c)^o]&=&\pi(a)[\omega,\pi(c)^o]\pi(b).
\eea
Thus, checking WC1 is equivalent to proving that the commutator of the elements of $G$ with $\pi(c)^o$ belongs to $(\Omega_F^1)^\ext$, for every anti-selfadjoint $c\in \A_F^\ext$. Moreover, if it is true for some $\omega\in G$, then it will be automatically true for its adjoint. Consider  a generic element 
\be
\pi(c)^o=[\mu\oplus m,\mu\oplus m,  q_\lambda,  q]\label{piczero}
\ee
with $c\in\A_F^\ext$ anti-selfadjoint. Then it is immediate that $[\omega_Y,\pi(c)^o]=0$, and  $[\omega_M,\pi(0,0,q,0)^o]=0$. Thus there only remains to compute $[\omega_M,\pi(0,\mu,0,m)^o]$ and $[\omega_M,\pi(\lambda,0,0,0)^o]$.

\subsection{Conditions on $M$ coming from the commutator of $\omega_M$ with $\pi(0,\mu,0,m)^o=[\mu\oplus m,\mu\oplus m,0,0]$}
First we can suppose without loss of generality that $\mu=0$ since   $0\oplus (m-\mu)$ is also anti-selfadjoint and has the same commutators as $\mu\oplus m$. One must have $[\omega_M,[0\oplus m,0\oplus m,0,0]]=\sum_i\pi(a_i)\omega_i\pi(b_i)$, with $a_i,b_i\in\A_F^\ext$ and $\omega_i\in G$. Focusing on the $(3,1)$-block we can discard all $\omega_i$ except $\omega_M$ and we obtain, with obvious notations:
\be
\colvec{M_{ll}&M_{lq}\cr M_{ql}&M_{qq}}\colvec{0&0&\cr 0&  m}=\sum_i\colvec{\mu_i&0\cr 0&  m_i}\colvec{M_{ll}&M_{lq}\cr M_{ql}&M_{qq}}\colvec{q_{\lambda_i}&0\cr 0&q_{\lambda_i}\otimes 1_3}
\ee
Since we already know from section \ref{SMback} that $M_{qq}=0$ we obtain the single condition
\be 
 M_{lq}  m = M_{lq}(\sum \mu_iq_{\lambda_i})\otimes 1_3 
\ee
After transposition and using $M_{ql}=M_{lq}^T$ this is put in the form

\be
  m^TM_{ql}=\delta(m)\otimes 1_3 M_{ql}
\ee
where $\delta(m):=\colvec{\alpha(m)&0\cr 0&\beta(m)}$ is some diagonal $2\times 2$ matrix depending on $m$, and $M_{ql}:=\colvec{X^1\cr X^2\cr X^3}$, with each $X^i\in\End(\CC^2)\otimes \End(\CC^3_g)$. Further decomposition with respect to $\CC^2$ yields
\bea
  m^TM_{ql}^{uu}=\alpha(m) M_{ql}^{uu},&&  m^TM_{ql}^{ud}=\alpha(m) M_{ql}^{ud}\cr
  m^TM_{ql}^{du}=\beta(m) M_{ql}^{du},&&  m^TM_{ql}^{dd}=\beta(m) M_{ql}^{dd}
\eea
The first equation shows that the $9$ components $(M_{ql}^{uu})_{jk}$ wrt. generations are $3\times 1$ vectors belonging to an eigenspace of $m^T$. Since this is true for any $m$, they must vanish, and we obtain $M_{ql}^{uu}=0$. The same is true of $M_{ql}^{ud},M_{ql}^{du}$ and $M_{ql}^{dd}$,  hence $M_{ql}=0$ and by symmetry $M_{lq}=M_{ql}^T=0$.

\subsection{Conditions on $M$ coming from the commutator with $\pi(\lambda,0,0,0)^o=[0,0,\tilde q_\lambda,0]$. }
It suffices to consider the case $\lambda=i$. We find that in order for $[\omega_M,\pi(i,0,0,0)^o]$ to belong to $(\Omega^1_F)^\ext$ there must exist $q_i,m_i,\lambda_i,\mu_i$ such that $  q_i M=\sum (\mu_i\oplus m_i)M  q_{\lambda_i}$, which yields
 
\be
q_i M_{ll}= M_{ll}\sum_i\mu_iq_{\lambda_i} 
\ee
Thus, we find
\be
\colvec{i&0\cr 0&-i  } \colvec{M_{ll}^{uu}&M_{ll}^{ud}\cr M_{ll}^{du}&0}= \colvec{M_{ll}^{uu}&M_{ll}^{ud}\cr M_{ll}^{du}&0}\colvec{\sum_i\mu_i\lambda_i&0\cr 0&\sum_i\mu_i\bar\lambda_i}
\ee
We obtain
\bea
i M_{ll}^{uu}&=&(\sum_i \mu_i\lambda_i)M_{ll}^{uu}\cr
i M_{ll}^{ud}&=&(\sum_i \mu_i\bar\lambda_i)M_{ll}^{ud}\cr
-i M_{ll}^{du}&=&(\sum_i \mu_i\lambda_i)M_{ll}^{du}
\eea
By the first and third equations, we find that $M_{ll}^{uu}\not=0\Rightarrow M_{ll}^{du}=0$. Hence $M_{ll}$ is either diagonal or off-diagonal in the $u/d$ indices.

\section{The general form of $D_F$ and its physical implications}
We have proven in sections \ref{SMback} and \ref{sectionWC1}    that in order for $\B_{\rm SM}$ to satisfy C1 and $\B_{\rm SM}^\ext$ to satisfy WC1, it is necessary and sufficient that the blocks of the Dirac operator satisfy:
\begin{itemize}
\item $Y$ is diagonal in the $l,q$ indices, with $Y_{qq}$ acting only on generations,
\item $M$ is diagonal in the $l,q$ indices with $M_{qq}=0$ and $M_{ll}$ is either of the form $\colvec{M_\nu&0\cr 0&0}$ with $M_\nu$ symmetric or $\colvec{0&M_{\nu e}\cr M_{\nu e}^T&0}$,
\item $B=0$.
\end{itemize} 
It is then an easy check that WC1' necessarily holds under these conditions, so that we have no more theoretical contraints on $D_F$. We now explore the physical consequences of this form for the Dirac. The diagonal solution for $M_{ll}$ provides a Majorana mass term for the right-handed neutrino, while the off-diagonal one yields an interaction between the right electron and  anti-neutrino mediated by a charged scalar field. Experimental evidence leads us to favour the former and there is no fine-tuning in doing so since we have only two distinct choices. However such a problem arises about $Y_{ll}$ and $Y_{qq}$: one would like them to be diagonal in the u/d indices in order to uniquely recover the Standard Model as a low energy limit, i.e. we want $Y$ to belong to a set of zero measure. It has been suggested \cite{connesmarcolli} to supplement the order 1 condition by the ``massless photon'' condition:
\be 
[D_F,\pi_F(\lambda,q_\lambda,0)]=0\label{masslessphotoncond}
\ee
for all $\lambda\in\CC$. This indeed removes the off-diagonal parts of $Y_{ll}$ and $Y_{qq}$ and ensures that the photon does not get a mass term. However  \eqref{masslessphotoncond} has no obvious generalization to spectral triples defined over  algebras different from $\A_F$, so that there isn't any theoretical motivation for this condition at the present moment. We should also mention that while \eqref{masslessphotoncond} is clearly sufficient to forbid a mass-term for the photon, it does not seem obvious that this condition is necessary. For these reasons, we believe that it is worthwhile to investigate the physical implications of generic off-diagonal terms in $Y_{ll}$ and $Y_{qq}$, which we rewrite more simply as $Y_\ell$ and $Y_q$, only to eventually discard them in front of any experimental  inconsistency, acknowledging a fine-tuning problem.

Let us  write $Y_{\ell/q}=Y_{\ell,q}^{\rm diag}+Y_{\ell,q}^{\rm off}$, with obvious meanings. The $1$-forms generated by $\omega_Y$ are then sums of terms like $\omega_{qYq_\lambda}$, and we have $qYq_\lambda=qq_\lambda Y^{\rm diag}+qq_{\bar \lambda}Y^{\rm off}$. If we write $q_1=qq_\lambda$ then $qq_{\bar \lambda}=q_1q_{e^{i\theta}}$, where $\theta$ is a phase which is independent from $q_1$. By summing $q_1Y^{\rm diag}+q_1e^{i\theta}Y^{\rm off}$ and $q_1Y^{\rm diag}+q_1e^{i(\theta+\pi)}Y^{\rm off}$ we can cancel the off-diagonal term, so that in the end $(\Omega^1_F)^\ext$ contains the $1$-forms $\omega_{q_1Y^{\rm diag}}$ and $\omega_{q_2Y^{\rm off}}$ where $q_1,q_2$ are independent quaternions.  The forms $\omega_{q_1Y^{\rm diag}}$ and $\omega_{q_2Y^{\rm off}}$ are themselves independent except if $Y^{\rm off}=\colvec{0&k\cr -\bar k&0}Y^{\rm diag}$. In this latter case we can write $q_1Y^{\rm diag}+q_2 Y^{\rm off}=(q_1+q_2q_k)Y^{\rm diag}$ and we obtain   the   field   $\Phi(q)$, where $q=q_1+q_2k$ is a generic quaternion. This case is thus similar to diagonal one\footnote{There is a  change in the value of the vev and in the relationship between the matrix $Y$ and the Dirac mass matrices. These two changes cancel each other as far as the prediction of the coupling constants are concerned, so that the analysis of \onlinecite{Bes-20} would remain unchanged.}.  If we are not in this particular case, we see that the scalar fields of theory are
\be 
\Theta(q,q',z):=\Phi(q)+\Phi(q)^o+\Phi'(q')+\Phi'(q')^o+\sigma(z)
\ee
with
\bea
\Phi(q)&=&\colvec{
0&-(Y^{\rm diag})^\dagger q^\dagger&0&0\cr
qY^{\rm diag}&0&0&0\cr
0&0&0&0\cr
0&0&0&0},\cr
\Phi'(q')&:=&\colvec{0&-(Y^{\rm off})^\dagger (q')^\dagger&0&0\cr
q'Y^{\rm off}&0&0&0\cr
0&0&0&0\cr
0&0&0&0},\cr
\sigma(z)&=&\colvec{0&0&-\bar zM^\dagger&0\cr
0&0&0&0\cr
zM&0&0&0\cr
0&0&0&0},
\eea
with $q,q'\in\HH$ and $z\in \CC$. We must write the scalar fields as fluctuations of the finite Dirac in order to compute the curvature, i.e. we write $\Theta(q,q',z)=D_F+\Theta(q-1,q'-1,z-1)$ and the scalar potential is 
\be
V(q,q',z)=\tr(P(\rho(\Theta(q-1,q'-1,z-1))^2)
\ee
Thus $V$ is by construction positive and vanishes for $q=q'=z=1$. The $\sigma$ field, which is anyway extremely massive\cite{Bes-20}, will not play any role in the sequel so that we can ignore it. The complete computation is very involved, but we can easily guess the form of the result. Indeed, we have
\bea
\rho(\Phi+\Phi'+\Phi^o+(\Phi')^o)&=&\rho(\Phi)+\rho(\Phi')+\rho(\Phi)^o+\rho(\Phi')^o+\{\Phi ,\Phi' \}+\{\Phi ,\Phi' \}^o 
\eea
and it is easy to see that all the pieces are orthogonal to each other. Hence the potential has the form
\be 
V(q,q')=F(q)^2+G(q')^2+H(q,q')^2\label{guess}
\ee
where $F,G,H$ are real functions which are gauge invariant. It follows that $F$ is a function of $|q|$, $G$ is a function of $|q'|$, and the interaction term is a function of $q^\dagger q',(q')^\dagger q$ and $|q||q'|$. When $q'=1$ this must reduce to the SM case, that is $F(q)=|q|^2-1$. When $q=1$ the situation is similar, so that in the end the potential has the form
\be 
V(q,q')=V_0(|q|^2-1)^2+V_0'(|q'|^2-1)^2+H(q,q')^2
\ee
from which we see that there are no other minima than $q=q'=1$ up to gauge transformations.

We thus have at first sight  a perfectly legitimate   2-Higgs model. However, it turns out that the two Higgs have opposite hypercharges. This can be seen by looking at the covariant derivatives, which are given by\cite{vs,blstandalone}:
\bea
D_\mu \Phi(q)&=&\partial_\mu \Phi(q)+[A_\mu,\Phi(q)],\cr
D_\mu \Phi'(q')&=&\partial_\mu \Phi'(q')+[A_\mu,\Phi'(q')]\label{covdev1}
\eea
where the gauge field $A_\mu$ is   an anti-selfadjoint field with values in $\pi_F(\A_F)$. Since $A_\mu^o$ commutes with $\Phi(q)$ and $\Phi'(q')$ we can replace $A_\mu$ with $A_\mu-A_\mu^o$, which is the usual Lie-algebra valued gauge field in the above formula. The commutators of $\Phi(q)$ and $\Phi'(q')$ with the Lie-algebra generators are exactly the same except the ones with the generator  of hypercharge which have opposite signs. Hence we can infer the computation of $D_\mu\Phi(q')$ from the one of $D_\mu\Phi(q)$ already performed in ref.  \onlinecite{blstandalone}. The kinetic terms come out as:
\bea
\tr(D_\mu\Phi(q)D^\mu\Phi(q))&=&-2a|D_\mu H|^2\cr
\tr(D_\mu\Phi'(q')D^\mu\Phi'(q'))&=&-2a|D_\mu H'|^2\label{covdev2}
\eea
where $a$ is a constant and $H,H'$ are the second columns of $q$ and $q'$ respectively: these will be the two complex doublets of the model when correctly normalized as $H_1=4\sqrt{a}H$ and $H_2=4\sqrt{a}H'$. We can deduce that the two vev's are the same since the minimum of the potential   is attained at $q=q'=1$ which corresponds to $H_1=H_2=\colvec{0\cr v}$, with $v=4\sqrt{a}$.  

If we expand the two fields around the vev's we can write
\be 
H_i=\colvec{0\cr v}+\ldots, i=1,2
\ee
where in the dots we hide $8$ real fields, only $3$ of them can be made to vanish by a gauge transformation. The mass terms for the gauge bosons then come from the $v^2$ term in
\be 
|D_\mu H_1|^2+|D_\mu H_2|^2
\ee
The covariant derivatives come out by direct computation from  \eqref{covdev1} as
\be 
D_\mu H_i=(\partial_\mu+ig_2T_3W_{\mu,a}\sigma^a+\frac{1}{2}iY_wg_1Y_\mu)H_i
\ee
where $T_3=\frac{1}{2}$ for both $H_1$ and $H_2$ and $Y_w=+1$ for $H_1$ and $-1$ for $H_2$. Let us rewrite this covariant derivatives in term of the fields which are usually mass eigenstates, namely
\bea
W_\mu&=&\frac{1}{\sqrt{2}}(W_\mu^1+iW_\mu^2),\cr
W_\mu^*&=&\frac{1}{\sqrt{2}}(W_\mu^1-iW_\mu^2),\cr
Z_\mu&=&c_w W_\mu^3-s_wY_\mu,\cr
A_\mu'&=&s_wW_\mu^3+c_wY_\mu,
\eea
where $c_w,s_w$ are the cosine and sine of the electroweak mixing angle and the photon field has been written $A_\mu'$ to distinguish it from the $A_\mu$ of \eqref{covdev1}. Then one finds
\be 
D_\mu=\partial_\mu+i \hat W+iA_\mu'\hat Q+iZ_\mu \hat L
\ee
where the operators $\hat W,\hat Q$ and $\hat L$ are
\bea
\hat W&=&T_3g_2\colvec{0&W_\mu^*\cr W_\mu&0},\cr
\hat Q&=&\frac{g_1g_2}{\sqrt{g_1^2+g_2^2}}(T_3\sigma^3+\frac{Y_w}{2}1_2),\cr
\hat L&=&\frac{1}{\sqrt{g_1^2+g_2^2}}(g_2^2T_3\sigma^3-g_1^2\frac{Y_w}{2}1_2)
\eea
Note that the  operators $\hat Q$ and $\hat L$ depend on the hypercharge and will not be the same when applied to $H_1$ or $H_2$. We will thus write them $\hat Q_{1,2}$ and $\hat L_{1,2}$ according to the case.

Since the vacuum state for both fields is $ve_{\downarrow}:=v\colvec{0\cr 1}$, one finds that the $v^2$ term in $|D_\mu H_1|^2+|D_\mu H_2|^2$ is 
\bea
M&:=&\bra ( \hat W+A_\mu'\hat Q_1+Z_\mu\hat L_1)e_{\downarrow}, (\hat W+A_\mu'\hat Q_1+Z_\mu\hat L_1)e_{\downarrow}\ket\cr
&&+\bra ( \hat W+A_\mu'\hat Q_2+Z_\mu\hat L_2)e_{\downarrow}, (\hat W+A_\mu'\hat Q_2+Z_\mu\hat L_2)e_{\downarrow}\ket
\eea
With $\hat W$   off-diagonal and $\hat Q_{1,2}$, $\hat L_{1,2}$   diagonal it is easy to see that the terms $\bra \hat W e_{\downarrow},\hat Q e_\downarrow\ket$ and $\bra \hat W e_{\downarrow},\hat L e_\downarrow\ket$ disappear. Now let us look at $\hat Q_{1,2}$, which is (proportional to) the electric charge operator. Since the $H_1$-field has hypercharge $+1$ we have $\hat Q_1\propto\colvec{1& 0\cr 0&0}$, so that $\hat Q_1 e_\downarrow=0$. On the other hand $\hat Q_2\propto\colvec{0& 0\cr 0&-1}$, so that $\hat Q_2 e_\downarrow\propto e_\downarrow$. In other words, in the vacuum state, the $H_1$-field is electrically neutral while the $H_2$-field has electric charge $-1$. The mass term $M$ is
\be 
M=2\bra \hat W^2e_\downarrow,e_\downarrow\ket+Z_\mu^2\bra (\hat L_1^2+\hat L_2^2)e_\downarrow,e_\downarrow\ket+(A_\mu')^2\bra \hat Q_2^2e_\downarrow,e_\downarrow\ket+2A_\mu'Z_\mu \bra \hat L_2\hat Q_2 e_\downarrow,e_\downarrow\ket
\ee
We see that the $W$-bosons are still mass eigenstates but the $Z$ and the photon are no longer. To find which mixing of the $Z$ and $A'$ have a definite mass we have to diagonalize the quadratic form given by the last $3$ terms of $M$, which can be computed to be, in matrix form, and up to a factor $1/4$:
\be 
q=\frac{1}{g_1^2+g_2^2}\colvec{2(g_1^4+g_2^4)& g_1g_2(g_2^2-g_1^2)\cr
g_1g_2(g_2^2-g_1^2)&g_1^2g_2^2}
\ee
Since $\det(q)=g_1^2g_2^2   \not=0$ there no longer exists a massless gauge boson in this theory, which for this reason has to be discarded.

\section{Beyond B-L ?}
We arrived at the conclusion that 
\be
M=\colvec{1&0\cr 0&0}\otimes \colvec{1&0_{1,3}\cr 0_{3,1}&0_{3,3}}\otimes M_\nu.\label{Mfac}
\ee
In this section we show that weak C1 makes $M_\nu$  vanish if we replace the algebra $\A_F^\ext$ with the Pati-Salam algebra 
\be 
\A_{\rm PS}=\HH\oplus \HH\oplus M_4(\CC)
\ee
represented on $\K_F$ by
\be 
\pi(p,q,m)=[p\otimes 1_4,q\otimes 1_4,1_2\otimes m,1_2\otimes m]\otimes 1_3.
\ee
In order to show this let us consider the $\bar R R$-block of a 1-form $\omega=\sum \pi(a_i')[D_F,\pi(a_i)]$, with $a_i,a_i'\in\A_{\rm PS}$. It is
\be 
\omega_{\bar R R}=\sum_i m_i'(Mp_i-m_iM)\label{decadix}
\ee
with $p_i,q_i,p_i',q_i'\in \HH$, $m_i,m_i'\in M_4(\CC)$. It follows that the general form of $\omega_{\bar R R}$ is
\be 
\omega_{\bar R R}=\colvec{\alpha&\beta\cr 0&0}\otimes \colvec{x&0\cr \vec{X}&0}\otimes M_\nu\label{formeomega31}
\ee
where $\alpha,\beta,x\in\CC$ and $\vec{X}\in\CC^3$.  Now if $a$ is an anti-selfadjoint element of $\A_{\rm PS}$, $\pi(a)^o$ is of the form $[g,g,p,q]$ with $g$ an anti-hermitian matrix and $p,q$ two pure quaternions, and the $\bar RR$ part of $[\omega,\pi(a)^o]$ is 
\be 
(\colvec{\alpha&\beta\cr 0&0}\otimes \colvec{x&0\cr \vec{X}&0}g-p\colvec{\alpha&\beta\cr 0&0}\otimes \colvec{x&0\cr \vec{X}&0})\otimes M_\nu \label{bleue}
\ee
which is not always of the form \eqref{formeomega31}, except if $M_\nu=0$. The problem is that the quaternions $p_i$ act on the right of $M$ in \eqref{decadix} wheras $p$ acts on the left in \eqref{bleue}. To keep the same form for the result the only possibility would be to act only with $p_i,q\in\CC\subset\HH$. Similarly the matrices $m_i,m_i'$ which act on the left in \eqref{decadix} and on the right in \eqref{bleue}, so that we have to restrict them to the subalgebra $\CC\oplus M_3(\CC)$ of $M_4(\CC)$ in order to keep a common form for $\omega_{\bar RR}$. Hence the largest algebra sitting between $\A_F$ and $\A_{\rm PS}$ which allows for a Majorana mass term and the weak order 1 condition is $\A_F^\ext$.

\begin{rem} Note that passing to $\A_{\rm PS}$ does not solve the fine-tuning problem for the matrix $Y$. Instead, as can be seen by computing the commutator of $\pi(0,0,m)^o$ with $\omega_Y$, weak C1 requires that $Y_\ell=Y_q$, that is the unification of the Yukawa couplings for leptons and quarks. 
\end{rem}

\section{Conclusion}
In this paper we have shown that  when we combine the first-order condition on the SM background with the weakened form of this condition on the B-L extended background, we arrive at a general form for the Dirac operator which predicts a Majorana mass term for the right-handed neutrino as well as a second Higgs doublet. This second Higgs leads to a mass for the photon and must be suppressed by setting to zero the off-diagonal terms in the Yukawa mass matrices $Y_\ell$ and $Y_q$. This is a fine-tuning problem similar to the strong CP problem, to which it bears an indirect relation since the vev of the second Higgs yields CP-violating terms in the quark sector of the fermionic action\cite{lee74} (and is the only source of such terms in this model). We have also shown that the weak order 1 condition destroys the Majorana mass term in the Pati-Salam model.

Though we set ourselves in the context of Lorentzian NGT, these conclusions should hold also for   Euclidean spectral NCG, since WC1 must also be satisfied in this case according to the algebraic background point of view. However, since the form of the scalar potential would be different from \eqref{guess}, further analysis would be required to dismiss the very unlikely possiblity that the vacuum states and the coupling constants conspire to allow for a massless gauge boson in this model. Anyway, such a happy coincidence would be destroyed by the renormalization group flow.

 
Anyhow, a new principle seems to be needed to eliminate the off-diagonal part of the matrices $Y_\ell$ and $Y_q$, and maybe a new extension not sitting inside the Pati-Salam model.  
 
\section{Acknowledgement}
We thank Christian Brouder for many fruitful discussions.
\bibliographystyle{unsrt}
\bibliography{../generalbib/SSTbiblio}
\end{document}